# One-Dimensional Frenkel and Wannier Excitons in Electric Fields: Stark Effect, Ionization, Polarizability and Electroabsorption

Thomas Garm Pedersen

*Department of Materials and Production, Aalborg University, DK-9220 Aalborg Øst, Denmark*
*Email: tgp@mp.aau.dk*

One-dimensional semiconductors are characterized by strongly bound excitons. Therefore, the Frenkel regime of excitons localized within a few unit cells is readily reached and traditional Wannier exciton models become inadequate. In the presence of strong electric fields, excitons are polarized and, in extreme cases, ionized. Such strong-field effects have previously been described analytically for Wannier excitons. In the present work, we show that analytical results can be extended to the more involved Frenkel case as well. Hence, by analytically solving the difference equation describing Frenkel excitons in electric fields, we derive close-form expressions for resonances providing Stark shifts and ionization rates. Moreover, closed-form results for exciton electroabsorption spectra and dynamic polarizability are obtained.

## 1. Introduction

Excitons dominate the optical response of one-dimensional semiconductors. This is due to reduced screening as well as increased electron-hole overlap in confined geometries [1,2]. Among the prominent examples of such systems are nanowires [3], nanotubes [4,5], nanoribbons [6,7] and conjugated polymers [8]. Subjecting such structures to strong electric fields leads to electroabsorption and ionization that are important tools for characterization and manipulation of excitons. Studies of the optical response of bulk semiconductors in electric fields date back to Franz [9] and Keldysh [10]. Also, electroabsorption by bulk excitons has been considered [1,2,11,12]. More recently, studies have been extended to one-dimensional materials, both with and without excitons [13-28].

Excitons can be classified as Frenkel or Wannier type according to their spatial localization [16,19,25,26]. Thus, Frenkel excitons localized within a few unit cells are distinguished from Wannier excitons that are delocalized over several unit cells, c.f. Fig. 1. The degree of localization reflects the exciton binding energy, which is particularly large in the Frenkel case. The two types of excitons therefore differ in their linear and nonlinear optical response [16,26]. In addition, their responses to external electric fields differ. Thus, the weakly bound Wannier excitons are expected to be more readily polarized and ionized in a long-axis electric field [23,25,26,29,30].

Regarding theoretical modelling, Wannier excitons are frequently described using effective-mass continuum models for their envelope function [9-12,17,18,20-23]. In contrast, Frenkel excitons are sensitive to the atomistic details of the unit cell [16,26]. Without simplification, this unfortunately implies complicated models without analytical solutions. The simpler

Wannier case has been solved in several instances including descriptions derived from the "one-dimensional" hydrogen problem [31-34] and non-integer dimensional generalizations [34]. Moreover, analytical solutions exist for continuum models incorporating static electric fields [20,23].

A minimal Frenkel exciton model retaining the essential deviation from continuum descriptions can be constructed by treating unit cells as discrete sites, between which carriers can hop. Such discrete lattice models capture the essence of Frenkel exciton physics and, moreover, approach the Wannier case in the limit of weak electron-hole interaction. Mathematically, discrete lattice models can be formulated as difference (rather than differential) equations. Accordingly, kinetic energy is described by discrete hopping between sites and second-order derivatives in the two models, respectively. In fact, the discrete model may be regarded as a discretized version of the continuum description. Thus, for states delocalized over many sites, the two models agree completely.

In the present work, we explore a minimal Frenkel exciton model including a static electric field directed along the lattice. This model was recently considered in Ref. [26], stating that no analytical solution was known. Here, we fill this gap by demonstrating that the model admits rather simple analytical solutions for eigenstates and energy resonances in arbitrary fields. This, then, provides a simple, closed-form expression for the excitonic optical response. Crucially, the energy resonances are complex-valued with imaginary parts describing field-induced ionization of the excitons. We investigate the ionization rate and demonstrate its suppression by strong exciton binding. Finally, a perturbative analysis of corrections due to electric fields yields explicit results for energy correction up to eighth order in the electric field. This also includes the second-order response to time-dependent fields and, thereby, the dynamic polarizability of Frenkel excitons. All calculations are contrasted by corresponding results for the Wannier limit, demonstrating excellent agreement in cases of weak electron-hole interaction.

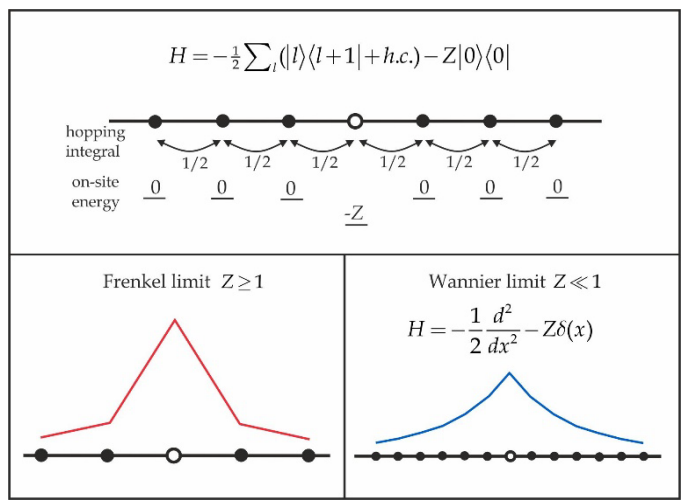

Figure 1. Discrete exciton model with hopping terms and an attractive on-site term at the origin. The lower panels illustrate Frenkel and Wannier limits, with the latter admitting a continuum description.

## 2. Frenkel and Wannier Models

The exciton model is formulated in terms of the relative electron-hole separation $l$, forming a discrete one-dimensional lattice with unit spacing, i.e. we take the lattice constant as our distance unit. In addition, we choose energy units, in which the nearest-neighbor hopping integral is (minus) one-half. The electron-hole attraction is taken to be short-ranged, i.e. restricted to a single unit cell and represented by an energy $-Z$ with $Z \geq 0$. This leads to the discrete model shown in Fig. 1 with *h.c.* implying the Hermetian conjugate. In Appendix A, a brief derivation of both Frenkel and Wannier models is given. When a static electric field $F \geq 0$ is added to the Frenkel model, eigenstates $\phi(l)$ obey the difference equation

$$E\phi(l) = F\phi(l)l - Z\phi(0)\delta_{l0} - \tfrac{1}{2}\{\phi(l+1) + \phi(l-1)\}, \tag{1}$$

with $E$ the associated energy eigenvalue. In addition to the lattice constant, the elementary charge is absorbed into the definition of $F$. The strategy for solving this problem is to consider the regions $l < 0$ and $l > 0$ and, subsequently, match at $l = 0$. In the absence of a field, the ground state energy (see Appendix B) is $E_0 = -\sqrt{1+Z^2}$ and the continuum of excited states spans the range $E \in [-1,1]$. In a positive field $F > 0$, the region $l > 0$ features an increasing potential energy from the field as $l$ increases. Hence, the boundary condition in this case becomes $\lim_{l \to \infty} \phi(l) = 0$. As long as $l \neq 0$, no effects of Coulomb interactions appear, and the exact solution with energy $E = -nF$ is a combination of Bessel functions of first and second kind with order varying with site index $l$,

$$\phi(l) = N \begin{cases} AJ_{l+n}(\tfrac{1}{F}) + BY_{l+n}(\tfrac{1}{F}) & l < 0 \\ J_{l+n}(\tfrac{1}{F}) & l > 0. \end{cases} \tag{2}$$

Note, that a $Y_{l+n}(\tfrac{1}{F})$ term is absent if $l > 0$ in order to comply with boundary conditions. Next, to find a general universal solution valid for arbitrary $Z$, we demand these both hold at $l = 0$, so that

$$AJ_n(\tfrac{1}{F}) + BY_n(\tfrac{1}{F}) = J_n(\tfrac{1}{F}). \tag{3}$$

Finally, considering Eq.(1) at $l = 0$, we see that

$$2(Z - nF)J_n(\tfrac{1}{F}) + AJ_{n-1}(\tfrac{1}{F}) + BY_{n-1}(\tfrac{1}{F}) + J_{n+1}(\tfrac{1}{F}) = 0. \tag{4}$$

Solving for $A$ and $B$, and introducing the step function $\Theta(-l)$, this means that

$$\phi(l) = N\left\{J_{l+n}(\tfrac{1}{F}) + \frac{\pi Z}{F} J_n(\tfrac{1}{F})\Theta(-l)\left[J_{l+n}(\tfrac{1}{F})Y_n(\tfrac{1}{F}) - Y_{l+n}(\tfrac{1}{F})J_n(\tfrac{1}{F})\right]\right\} \tag{5}$$

with normalization (in the delta-function sense)

$$N = \frac{\sqrt{F}}{\sqrt{(\pi Z)^2 J_n^4(\frac{1}{F}) + \{F + \pi Z J_n(\frac{1}{F}) Y_n(\frac{1}{F})\}^2}}. \qquad (6)$$

We will now contrast this with the continuum model. The discrete model applies universally regardless of the value of $Z$. Yet, a simplified model emerges in the Wannier limit $Z \ll 1$. In this case, the difference equation Eq.(1) can be replaced by the more conventional differential one. Hence, the discrete coordinate $l$ is replaced by a continuous $x$ with associated wave function $\varphi(x)$. The discrete interaction $-Z|0\rangle\langle 0|$ then becomes a Dirac delta function $-Z\delta(x)$ such that (shifting the energy by one unit),

$$\left\{-\frac{1}{2}\frac{d^2}{dx^2} - Z\delta(x) + Fx\right\}\varphi(x) = E\varphi(x). \qquad (7)$$

As detailed in previous work [23,30,32,33], the delta potential implies a discontinuous derivative at $x=0$. In the presence of a field, the general solution is a linear combination of Airy function Ai and Bi. Essentially, these functions replace the Bessel functions $J$ and $Y$, respectively. The divergent behavior of Bi means that only the Ai solution can be used for $x>0$. For $x<0$, both solutions oscillate and any linear combination is, in principle, acceptable. That is, with $\mu = (2F)^{1/3}$

$$\varphi(x) = N \begin{cases} A\mathrm{Ai}[\mu(x+n)] + B\mathrm{Bi}[\mu(x+n)] & x<0 \\ \mathrm{Ai}[\mu(x+n)] & x>0. \end{cases} \qquad (8)$$

Boundary conditions for function and derivative at $x=0$ eventually imply that [23]

$$\varphi(x) = N\left\{\mathrm{Ai}[\mu(x+n)] + \Theta(-x)\frac{2\pi Z}{\mu}\mathrm{Ai}(\mu n)[\mathrm{Ai}(\mu n)\mathrm{Bi}[\mu(x+n)] - \mathrm{Bi}(\mu n)\mathrm{Ai}[\mu(x+n)]]\right\} \qquad (9)$$

with

$$N = \frac{\sqrt{\frac{1}{2}\mu}}{\sqrt{(\pi Z)^2 \mathrm{Ai}^4(\mu n) + \{\frac{1}{2}\mu - \pi Z \mathrm{Ai}(\mu n)\mathrm{Bi}(\mu n)\}^2}}. \qquad (10)$$

Note the striking similarity between Eq.(6) and Eq.(10). The close resemblance between the discrete and continuum solutions is further evidenced in Fig. 2 illustrating the two wave functions in the case $Z=0.1$, for which the Wannier approximation is expected to be reasonable. Excellent quantitative agreement between the two solutions is observed, in particular, in the vicinity of the origin. Physically, the origin represents the amplitude for electron and hole locations to coincide, which is directly linked to the optical response, as discussed below. Thus, we expect response functions in the two models to agree as well.

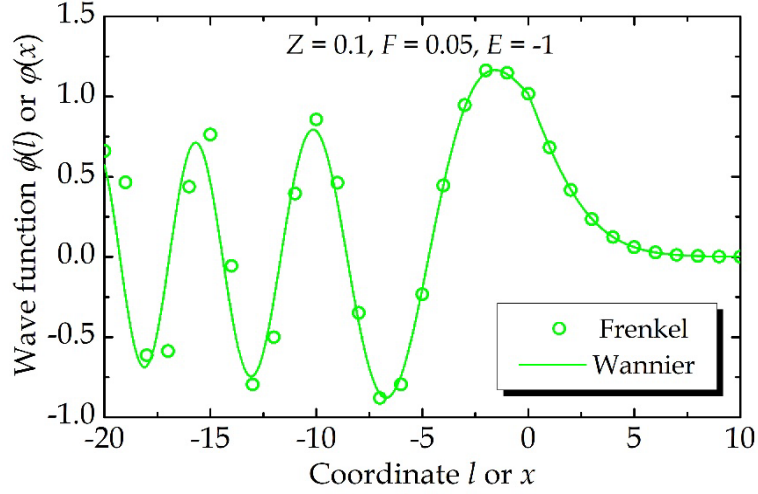

Figure 2. Illustrative example of wave functions pertaining to Frenkel (circles) and Wannier (solid curve) models. Parameters ($Z=0.1, F=0.05, E=-1$) represent a case of relatively weak electron-hole interaction, for which the two models should nearly agree.

The optical response is of central importance in exciton physics. As shown in Appendix A, the excitonic transition from a filled valence band is described by a momentum matrix element $P \approx p_{cv}\phi(0)$, where $p_{cv}$ is the constant Bloch-function part. It follows that the absorptive (imaginary) interband optical response at a frequency $\omega$ can be approximated by $\sim |p_{cv}|^2 \chi(\omega)$, where $\chi(\omega) = |\varphi_{E=\omega}(0)|^2$ or $|\phi_{E=\omega}(0)|^2$ for Wannier and Frenkel models, respectively. In the Frenkel case, Eqs.(5-6) demonstrate that

$$\chi(\omega) = \frac{F J_{-\omega/F}^2(\tfrac{1}{F})}{(\pi Z)^2 J_{-\omega/F}^4(\tfrac{1}{F}) + \left\{F + \pi Z J_{-\omega/F}(\tfrac{1}{F}) Y_{-\omega/F}(\tfrac{1}{F})\right\}^2}. \qquad (11)$$

Similarly, in the Wannier model,

$$\chi(\omega) = \frac{\tfrac{1}{2}\mu \mathrm{Ai}^2(-\tfrac{\mu\omega}{F})}{(\pi Z)^2 \mathrm{Ai}^4(-\tfrac{\mu\omega}{F}) + \left\{\tfrac{1}{2}\mu - \pi Z \mathrm{Ai}(-\tfrac{\mu\omega}{F})\mathrm{Bi}(-\tfrac{\mu\omega}{F})\right\}^2}. \qquad (12)$$

Physically, the frequency $\omega$ should be taken relative to the semiconductor band gap $E_g$. Moreover, the different energy zero points applied in Eqs.(1) and (7) must be accounted for by shifting Wannier model spectra by one frequency unit.

The comparison between optical spectra is made in Fig. 3. Here, cases of strong ($Z=1$), medium ($Z=0.5$) and weak ($Z=0.1$) electron-hole interaction are considered. A phenomenological line-width of 0.05 is added by convolution with a Lorentzian. In the weakly coupled case $Z=0.1$, Frenkel and Wannier exciton spectra agree rather well in the vicinity of the band gap. However, marked discrepancies appear at higher energies. This is caused by the different nature of continuum states in the two models. Thus, in the absence of an electric field, the Frenkel model continuum has a finite width of two units, while the Wannier continuum is free-electron-like and has no upper bound. Therefore, in weak fields, the Frenkel spectrum is effective cut off at $\omega - E_g \approx 2$, whereas the Wannier absorption persists at high energies. This is a consequence of the finite band width with a large density of states at the band edge. Yet, the Wannier approximation remains good at lower energies even for Coulomb charges as large as $Z=0.5$. As $Z$ is increased, oscillator strength is transferred to the bound exciton. However, the Wannier model tends to overestimate both binding energy and absorption intensity, as clearly seen for $Z=1$.

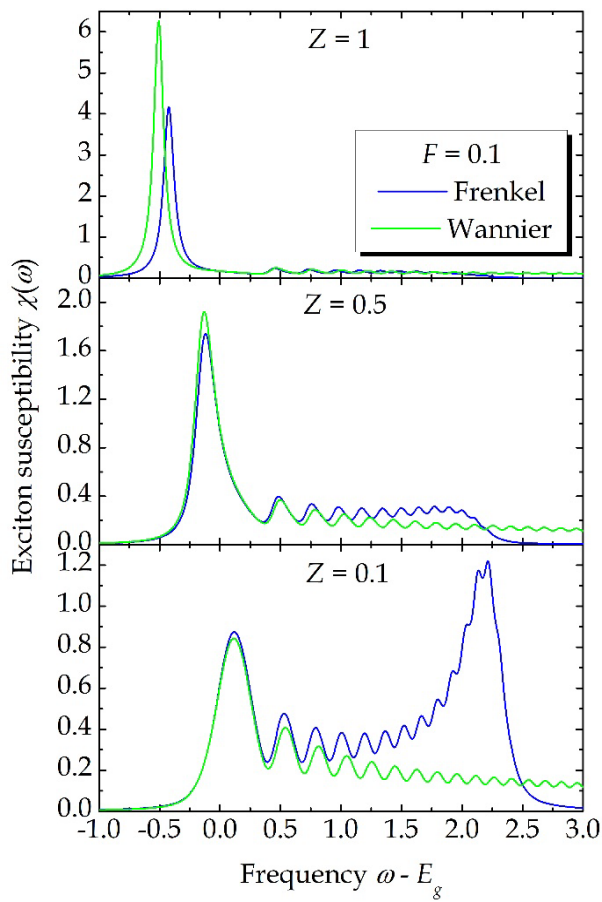

Figure 3. Absorption spectra of Frenkel (blue) and Wannier (green) excitons at a fixed electric field of $F=0.1$. The three panels represent cases of strong (top), medium (middle) and weak (bottom) electron-hole interaction. A phenomenological line-width of 0.05 is included in all spectra.

## 3. Exciton Ionization and Polarizability

In large electric fields, the probability of field-induced exciton ionization rises markedly. Ionization itself is a highly non-perturbative phenomenon that cannot be captured by perturbation expansion at any finite order. Analytical solutions including the field in a non-perturbative manner are rare, however, and necessarily restricted to rather simple models [23]. As we will demonstrate, both the discrete Frenkel and continuum Wannier model allow for such analytical results.

We seek an analytical resonance condition having field-dependent resonances as solutions. Such resonances can be considered complex-valued energy eigenvalues with an imaginary part providing the ionization rate $\Gamma = -2\,\text{Im}\,E(F)$. Similarly, the real part provides the Stark-shifted energy to all orders in the electric field. Now, both Eq.(5) and Eq.(9) are derived under the standard condition of real-valued energies consistent with delta function normalization. In contrast, complex resonances result from outgoing-wave boundary conditions. Hence, we seek a condition for the "energy" $E$ such that Eqs.(2) and (8) approach an outgoing wave to the far left $l, x \to -\infty$ asymptotically. Applying asymptotic forms of both Bessel and Airy functions, it is found that the ratio $B/A$ must equal $i$ and $-i$ in Eq.(2) and Eq.(8), respectively. In turn, this implies the resonance conditions

$$ZJ_n(\tfrac{1}{F})\{J_n(\tfrac{1}{F}) + iY_n(\tfrac{1}{F})\} + \frac{i}{\pi}F = 0 \tag{13}$$

in the Frenkel case, and

$$\pi Z \text{Ai}(\mu n)\{\text{Ai}(\mu n) - i\text{Bi}(\mu n)\} + \tfrac{i}{2}\mu = 0 \tag{14}$$

in the Wannier case. In both cases, we note that these conditions correspond to diverging normalization constants, c.f. Eq.(6) and (10). Also, the mathematical structure of these conditions is remarkably similar.

Solving Eqs.(13-14) numerically provides Stark shifts $\Delta E = \text{Re}\,E(F) - E(0)$ and ionization rates $\Gamma = -2\,\text{Im}\,E(F)$. In Fig. 4, numerical results obtained for a range of Coulomb charges $Z$ are illustrated. Note that both electric fields and resonances have been scaled appropriately in order to enable comparison between different $Z$-values in a single plot. Several features are noteworthy in this plot. First, the Stark shift is initially negative, decreasing quadratically with field. At a characteristic field $\sim Z^2/2$, though, the Stark shift goes through a minimum and then increases monotonically. This behavior is similar to that of three- and low-dimensional hydrogen atoms [34]. At the same time, the ionization rate increases from an initially vanishing value to a finite rate, albeit exponentially small in small fields. Again, beyond a characteristic field of $\sim Z^2/10$ a marked increase in $\Gamma$ sets in. This is the range of significant field-induced ionization and, moreover, consistent with hydrogen-atom results [34]. Finally, it is seen that Frenkel and Wannier models are in excellent agreement for both real and imaginary parts of the resonance in the weakly interacting case $Z = 0.1$.

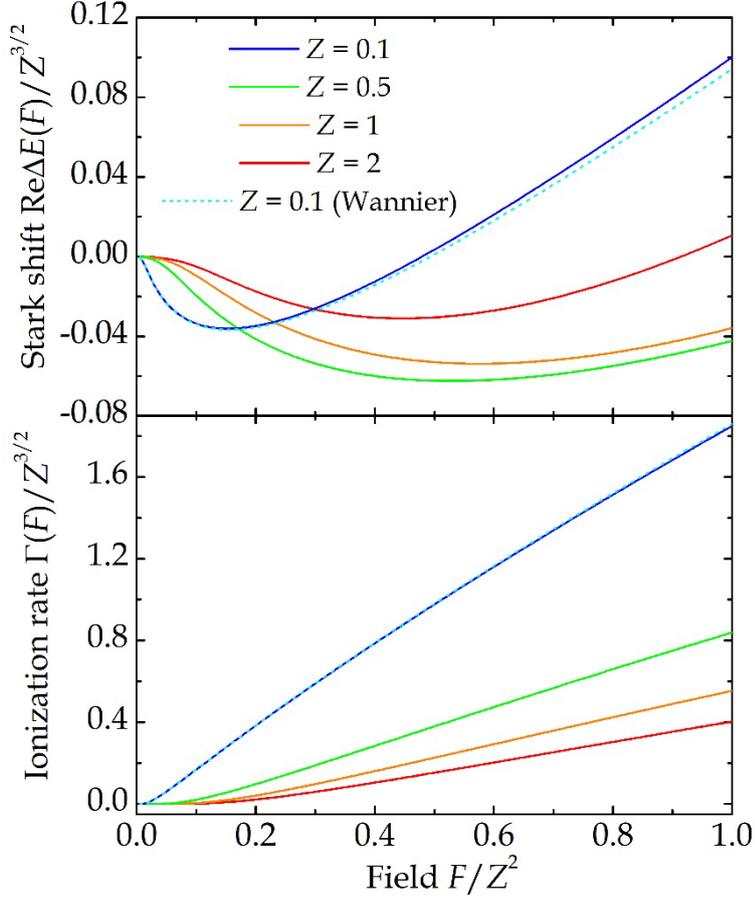

Figure 4. Stark shift (top panel) and ionization rate (lower panel) for a range of Z-values as shown by color. In addition, Wannier approximation results are shown for $Z = 0.1$ as dashed cyan curves.

In weak electric fields, a perturbative treatment of Stark shifts is expected to be adequate. In the Appendix B, Dalgarno-Lewis perturbation theory [35] is applied to derive exact energy corrections up to eighth order. This is achieved for both Frenkel and Wannier models as shown in Eqs.(B4-B7) and Eq.(B11), respectively. The energy series $E(F) = \sum_{p=0}^{\infty} E_{2p} F^{2p}$ is diverging badly, however. In fact, this is an asymptotic, rather than Taylor, series. Moreover, the radius of convergence is zero. This is particularly clear from the Wannier coefficients Eq.(B11) that agree with the low-Z limit of the Frenkel coefficients Eqs.(B4-B7). Yet, the Frenkel series diverges even at larger Z-values, albeit less badly. In the limit of very large Z, the Frenkel coefficients become $E_{2p} \approx -1/(2Z^{2p+1})$ that implies a radius of convergence equal to Z. Yet, for any finite Z, the radius of convergence vanishes, as evidenced by the finite imaginary parts of resonances derived from Eq.(13).

The divergent behavior is clearly visible in Fig. 5. Here, partially summed series of order 2 to 8 are compared to essentially exact results obtained by solving Eq.(13) numerically. Even at the relatively large value $Z=1$, including more terms in the perturbation series leads to more wildly diverging curves. It is therefore concluded that any partially summed perturbation series can be applied in a restricted range of weak fields only. Yet, there are means of making sense of such asymptotic series, as demonstrated in several instances [34,36,37]. A successful approach is hypergeometric resummation, in which a partial Taylor series is used to parameterize a hypergeometric function, such that the low orders agree [34,36,37]. Resummation based on $_2F_1$ hypergeometric functions, see Eq.(B12), requires five pieces of information for parameterization, which fits nicely with the eighth order perturbative series derived in the present work.

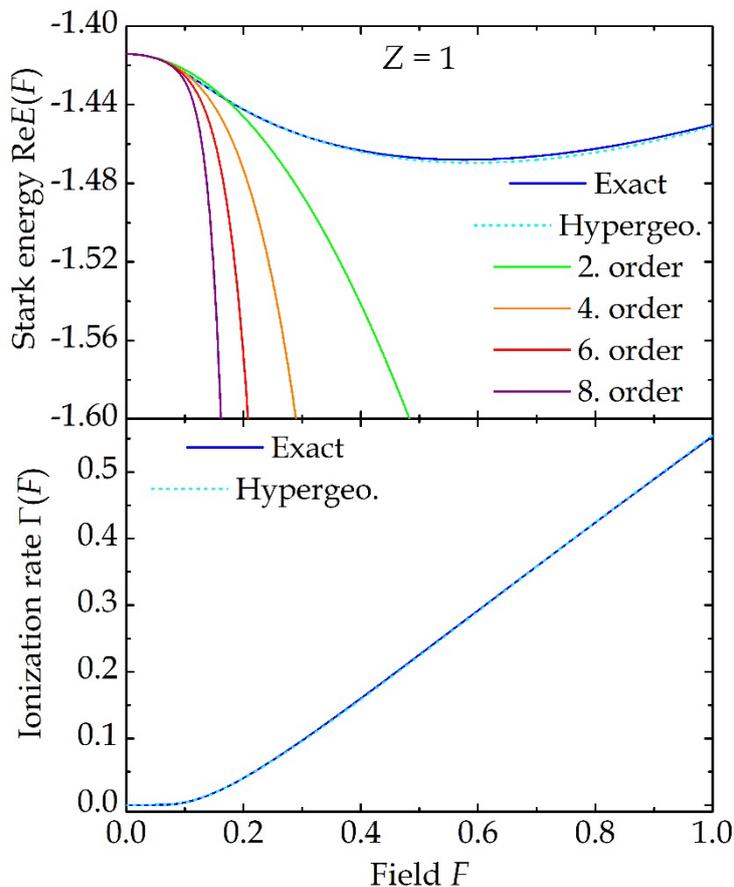

Figure 5. Comparison of field-dependent resonances in a strongly interacting ($Z=1$) case. Solid blue curves are numerically exact results, while other colors show truncated perturbation series. Dashed curves are results from hypergeometric approximants.

In Fig. 5, we include results calculated from a $_2F_1$ hypergeometric approximant as introduced in Ref. [34]. We select the shifted form Eq.(B12) capable of producing an imaginary part at arbitrarily small electric fields. As is clear from the figure, both real and imaginary parts of the resonance agree rather well with numerically precise values. This testifies to the usefulness of low-order perturbative information when combined with hypergeometric resummation allowing for reliable prediction of complex resonances in finite fields. Crucially, highly non-perturbative phenomena such as ionization rates and non-monotonic Stark shifts are accurately captured. Also, we have verified that similar accuracy is obtained for other Coulomb charges including weak interaction $Z \ll 1$.

In a weak electric field, the dominant energy correction $E_2$ is related to the static polarizability $\alpha(0)$ via $\alpha(0) = -2E_2$. However, a particular strength of Dalgarno-Lewis perturbation theory (in addition to exact energy corrections) is the ability [37,38] to yield dynamic polarizabilities $\alpha(\omega)$ appearing in response to monochromatic electric fields oscillating at frequency $\omega$. This response should not be confused with the dynamic *interband* response discussed above and depicted in Fig. 3. Rather, it describes the dynamic polarization of an exciton already created by an interband absorption event. Thus, the frequency in this case is understood to be relatively low corresponding to energies near the exciton binding energy.

A time-dependent field naturally requires solving the time-dependent Frenkel/Wannier equation. Writing the first-order correction to the wave function as $\phi_1(l,t) = \frac{1}{2}\{\phi_1^+(l)e^{-i\omega t} + \phi_1^-(l)e^{i\omega t}\}$ with $\phi_1^\pm(l) = \text{sign}(l)\psi_1^\pm(|l|)$, it can be shown that

$$\psi_1^\pm(l) = \frac{\sqrt{Z}}{\omega^2(1+Z^2)^{1/4}} \left\{ \frac{Z \pm \omega l}{(\sqrt{1+Z^2}+Z)^l} - \frac{Z}{\left(\sqrt{Z^2+\omega^2 \mp 2\omega(1+Z^2)^{1/2}} + \sqrt{1+Z^2} \mp \omega\right)^l} \right\}. \quad (15)$$

The dynamic polarizability $\alpha(\omega) = \langle \phi_0 | l | \phi_1^+ \rangle + \langle \phi_1^- | l | \phi_0 \rangle$ becomes

$$\alpha(\omega) = -\frac{1}{\omega^2\sqrt{1+Z^2}} + \frac{2Z^4}{\omega^4\sqrt{1+Z^2}} \left\{ 1 - \frac{\sqrt{\omega^2+Z^2+\sqrt{(Z^2-\omega^2)^2-4\omega^2}}}{\sqrt{2}Z} \right\}$$

$$= \frac{5^2+4Z^2}{4Z^4\sqrt{1+Z^2}} + \frac{21+28Z^2+8Z^4}{8Z^8\sqrt{1+Z^2}}\omega^2 + \frac{429+792Z^2+432Z^4+64Z^6}{64Z^{12}\sqrt{1+Z^2}}\omega^4 + O(\omega^6). \quad (16)$$

A similar calculation for the Wannier model [30,38] yields a dynamic polarizability

$$\alpha(\omega) = -\frac{1}{\omega^2} + \frac{Z^4}{\omega^4}\left\{2 - \sqrt{1-\frac{2\omega}{Z^2}} - \sqrt{1+\frac{2\omega}{Z^2}}\right\}$$

$$= \frac{5^2}{4Z^4} + \frac{21}{8Z^8}\omega^2 + \frac{429}{64Z^{12}}\omega^4 + O(\omega^6). \quad (17)$$

The dynamic polarizabilities of the two models are compared in Fig. 6. Again, we used scaled quantities to collect results for a wide range of Z-values in a single plot. The peaks seen in the real parts (and thresholds in imaginary parts) coincide with the exciton binding energy $\omega = \sqrt{1+Z^2} - 1$. As expected, the Wannier and Frenkel models are in excellent agreement at weak interaction $Z = 0.1$.

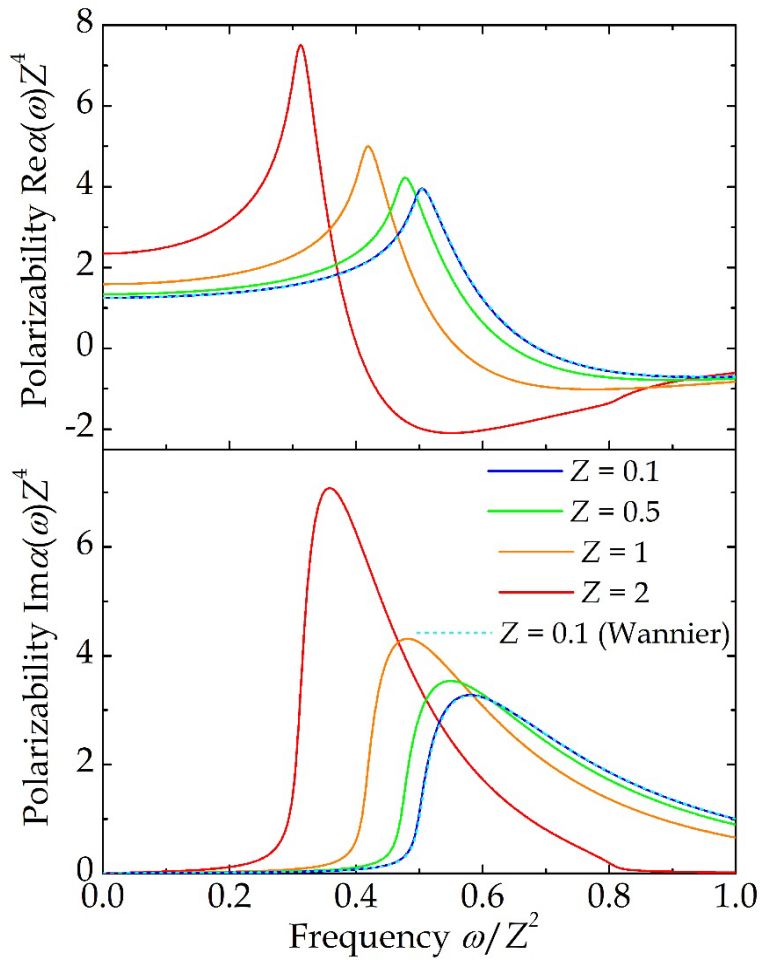

Figure 6. Real (top) and imaginary (bottom) parts of the dynamic polarizability for various values of the Coulomb charge, as distinguished by color. Dashed curves show Wannier equation results.

## 4. Summary

Summarizing, an analytical solution to the discrete eigenvalue problem describing Frenkel excitons in electric fields has been found. The solution is valid for arbitrary energy, field and Coulomb charge. Throughout, results for Frenkel excitons have been compared to the simpler Wannier exciton model. It is clearly demonstrated that the two agree in cases of weak electron-hole interaction. An analytical condition for complex energy resonances has been found and compared to perturbation series and hypergeometric resummation. In addition, an analytic expression for the dynamic polarizability is derived. Finally, the excitonic interband optical response is studied, highlighting the interplay between electric fields and Coulomb binding.

## Appendix A: Exciton Models

In this appendix, we briefly discuss Frenkel and Wannier exciton models. The starting point is the picture of two-band excitons as *k*-space superpositions $|exc\rangle = \sum_k \psi(k)|v_k \to c_k\rangle$, where $|v_k \to c_k\rangle$ is a singly-excited Slater determinant built by replacing an occupied valence band single-particle state $v_k$ by an empty conduction band state $c_k$ at identical *k*-points *k*. Neglecting exchange, the matrix equation for $\psi(k)$ is

$$E\psi(k) = E_{cv}(k)\psi(k) + \frac{1}{N}\sum_{k'}\int e^{i(k'-k)x}V(x)dx\psi(k'). \tag{A1}$$

Here, $E_{cv}(k) = E_c(k) - E_v(k)$ is the difference between conduction and valence band single-particle energies, $V(x)$ is the screened electron-hole Coulomb interaction and *N* is the number of *k* points. We also introduce the real-space wave function $\phi(x) = N^{-1}\sum_k e^{ikx}\psi(k)$. At this point, the Frenkel and Wannier approaches depart. Thus, in the former case, $E_{cv}$ is taken in the nearest-neighbor hopping form appropriate for a discrete lattice $E_{cv}(k) = -\cos(k) = -\frac{1}{2}(e^{ik} + e^{-ik})$, while in the latter, the effective mass form $E_{cv}(k) = \frac{1}{2}k^2$ is applied (recalling that lattice constants and hopping integrals are absorbed into the units). It follows that, applying the inverse Fourier transform to Eq.(A1) in the Frenkel and Wannier cases, we find

$$E\phi(x) = -\tfrac{1}{2}\{\phi(x+1) + \phi(x-1)\} + V(x)\phi(x) \quad \text{(Frenkel)}. \tag{A2}$$

and

$$E\phi(x) = -\tfrac{1}{2}\phi''(x) + V(x)\phi(x) \quad \text{(Wannier)}. \tag{A3}$$

Finally, approximating *V* by a contact interaction amounts to the projection $V(x) \to -Z|0\rangle\langle 0|$ in the Frenkel case, while $V(x) = -Z\delta(x)$ for Wannier excitons. Including the electric field eventually implies Eqs.(1) and (7). The interband optical response is derived from the momentum matrix element $P = \langle g|\sum_i p_i|exc\rangle$, where $|g\rangle$ is the completely filled valence band constituting the ground state Slater determinant, $p_i$ is the momentum operator of the *i*´th electron and we sum over all electrons. Evaluating matrix elements between Slater determinants, we then find

$$P = N^{-1}\sum_k \psi(k)\langle v_k|p|c_k\rangle. \tag{A4}$$

If the *k*-dependence of the band-to-band matrix element is ignored $\langle v_k|p|c_k\rangle \approx p_{cv}$ with $p_{cv}$ the value at the band gap, we finally arrive at $P \approx p_{cv}\phi(0)$.

## Appendix B: Dalgarno-Lewis Perturbation Theory

In this appendix, we treat the electric field perturbatively in order to obtain low-order Taylor series for the energy correction. Rather than summing over states, as in standard perturbation approaches, we apply the Dalgarno-Lewis method [35] that allows for exact closed-form results. The method proceeds by solving a sequence of perturbation problems obtained by collecting terms of a particular order. To this end, we expand

$$E = E_0 + E_2 F^2 + E_4 F^4 + ..., \quad \phi(l) = \phi_0(l) + F\phi_1(l) + F^2 \phi_2(l) + ... \quad (B1)$$

Note that no odd orders appear in the energy expansion due to inversion symmetry of the unperturbed system. Similarly, parity dictates that for the spatially symmetric ground state $\phi_n(l) = \text{sign}^n(l)\psi_n(|l|)$. Solutions may be expressed in terms of $\alpha \equiv \sinh^{-1} Z$, such that at zeroth order

$$\psi_0(l) = \sqrt{\tanh \alpha}\, e^{-\alpha l} = \frac{\sqrt{Z}}{(1+Z^2)^{1/4}(Z+\sqrt{1+Z^2})^l}, \quad (B2)$$

with energy $E_0 = -\cosh\alpha = -\sqrt{1+Z^2}$. Collecting first order terms in Eq.(B1) and solving, we get

$$\psi_1(l) = -l\frac{\cosh\alpha + l\sinh\alpha}{2\sinh^2\alpha}\sqrt{\tanh\alpha}\,e^{-\alpha l} = -\frac{l(lZ+\sqrt{1+Z^2})}{2Z^{3/2}(1+Z^2)^{1/4}(Z+\sqrt{1+Z^2})^l}. \quad (B3)$$

From these states, we get the second-order energy $E_2 = \langle \phi_0 | l | \phi_1 \rangle$ and proceeding in this way, we eventually find

$$E_2 = -\frac{5^2 + 4Z^2}{8Z^4\sqrt{1+Z^2}}, \quad (B4)$$

$$E_4 = -\frac{880 + 2343Z^2 + 2120Z^4 + 720Z^6 + 64Z^8}{128Z^{10}(1+Z^2)^{3/2}}, \quad (B5)$$

$$E_6 = -\frac{1}{1024Z^{16}(1+Z^2)^{5/2}}\Big\{340000 + 1512720Z^2 + 2710781Z^4 + 2482876Z^6 \\ + 1212080Z^8 + 295552Z^{10} + 28800Z^{12} + 512Z^{14}\Big\}, \quad (B6)$$

$$E_8 = -\frac{1}{32768 Z^{22}(1+Z^2)}\{1104000000 + 6885701760 Z^2 + 18476804256 Z^4$$
$$+ 27808204875 Z^6 + 25633434000 Z^8 + 14815994592 Z^{10} + 5269057152 Z^{12} \quad \text{(B7)}$$
$$+ 1079724288 Z^{14} + 110315520 Z^{16} + 4001792 Z^{18} + 16384 Z^{20}\}.$$

The expressions for the wave function corrections are rather involved and, therefore, omitted here.

In Fig. 7, the Z-dependence of all terms is illustrated applying logarithmic scale due to the pronounced variation with Z. Even with the dominant low-Z dependence scaled away, as in the right panel, the coefficients vary over several orders of magnitude when Z is varied from low to high values. This testifies to the pronounced difference between Frenkel and Wannier excitons. Specifically, as seen in the left-hand panel, all energy corrections diminish dramatically with Z. Hence, tightly bound Frenkel excitons are much less affected by electric fields, as expected. Also, the deviation from Wannier exciton theory (see Eq.(B11) below) is seen to set in around $Z \sim 1$ such that, above this value, Wannier models do not apply.

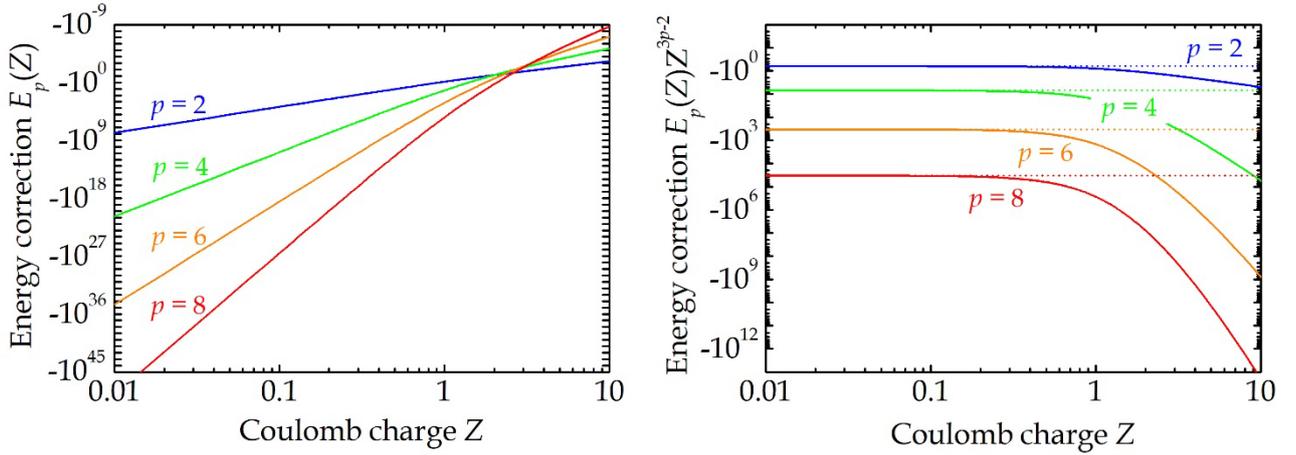

Figure 7. Coefficients of the eighth order Taylor expansion versus Z. In the right panel, the dominant Low-Z behavior is scaled away. Also, the Wannier approximation results are shown as horizontal dashed lines.

A perfectly analogous Dalgarno-Lewis calculation can be made for the Wannier equation. In this case, the first-order correction to the wave function satisfies the inhomogeneous Dalgarno-Lewis equation

$$\left\{-\frac{1}{2}\frac{d^2}{dx^2} - Z\delta(x) + \frac{Z^2}{2}\right\}\varphi_1(x) = -x\varphi_0(x), \quad \text{(B8)}$$

having the solution

$$\varphi_1(x) = -\frac{x}{2Z^2}(1+Z|x|)\varphi_0(x). \tag{B9}$$

In general, $\varphi_n(x) = \text{sign}^n(x)\psi_n(|x|)\varphi_0(x)$, so that

$$\begin{aligned}
\psi_1(x) &= -x\frac{1+xZ}{2Z^2}, \\
\psi_2(x) &= \frac{-30 + 15x^2Z^2 + 10x^3Z^3 + 3x^4Z^4}{24Z^6}, \\
\psi_3(x) &= -x\frac{30 + 30xZ + 45x^2Z^2 + 25x^3Z^3 + 7x^4Z^4 + x^5Z^5}{48Z^8}, \\
\psi_4(x) &= \frac{1}{1152Z^{12}}\{-40500 + 7020x^2Z^2 + 4680x^3Z^3 + 2295x^4Z^4 \\
&\quad + 876x^5Z^5 + 226x^6Z^6 + 36x^7Z^7 + 3x^8Z^8\}.
\end{aligned} \tag{B10}$$

For the energy corrections, we get

$$E_2 = -\frac{5}{8Z^4}, \quad E_4 = -\frac{55}{8Z^{10}}, \quad E_6 = -\frac{10625}{32Z^{16}}, \quad E_8 = -\frac{1078125}{32Z^{22}}. \tag{B11}$$

Note that these agree completely we the low-Z limits of Eqs.(B4-B7).

The eighth order Taylor series based on Eqs.(B4-B7) or Eq.(B11) can be resummed using the hypergeometric ansatz [34]

$$E(F) = E_0\left\{1 + h_4 F^2 \frac{\Gamma(h_0+h_1)\Gamma(l+h_2)}{\Gamma(h_0+h_1+h_2)} {}_2F_1(h_1, h_2, h_1+h_2+h_0, 1+h_3 F^2)\right\}, \tag{B12}$$

Here, $h_0$ is an auxiliary integer taken as $h_0 = 10$ and the remaining parameters $h_{1-4}$ are found by matching the Taylor expansion of Eq.(B12) to the known eighth order results. The rationale behind Eq.(B12), as compared to other hypergeometric approximants, is that a finite imaginary part is obtained in arbitrarily small fields. This is a consequence of the shifted argument $1 + h_3 F^2$ since the branch-cut of ${}_2F_1(h_1, h_2, h_3, z)$ is the line $z \in [1, \infty]$. The shifted argument enforces the additional integer $h_0 > 4$, however, in order for Eq.(B12) to yield a Taylor expansion without logarithmic terms.

## Acknowledgements

This work is supported by the DNRF Centre CLASSIQUE sponsored by the Danish National Research Foundation, grant nr. 187.